\documentclass[preprint, aps, nofootinbib]{revtex4}
\usepackage{CJK}
\usepackage{graphicx}
\usepackage{amsmath}
\usepackage{amsfonts}
\usepackage{amssymb}
\usepackage{color}%
\usepackage{dcolumn}
\usepackage{slashed}
\usepackage{multirow}

\providecommand{\U}[1]{\protect\rule{.1in}{.1in}}

\newcommand{\f}{\begin{equation}}
\newcommand{\ff}{\end{equation}}
\newcommand{\fa}{\begin{eqnarray}}
\newcommand{\ffa}{\end{eqnarray}}

\begin{document}
\title{Fermionic phase transition induced by the effective impurity in holography}
\author{Li-Qing Fang$^{1,2}$}
\email{fangliqing@sjtu.edu.cn}
\author{Xiao-Mei Kuang$^{3,4}$}
\email{xmeikuang@gmail.com}
\author{Bin Wang$^{1}$}
\email{wang_b@sjtu.edu.cn}
\author{Jian-Pin Wu$^{5,6}$}
\email{jianpinwu@mail.bnu.edu.cn}
\affiliation{${}^{1}$IFSA Collaborative Innovation Center, Department of Physics and Astronomy,
Shanghai Jiao Tong University, Shanghai 200240, China.\\
${}^{2}$School of Physics and Electronic Information, Shangrao Normal University, Shangrao 334000, China.\\
${}^{3}$Department of Physics, National Technical University of Athens, GR-15780 Athens, Greece.\\
${}^{4}$Instituto de F\'isica, Pontificia Universidad Cat\'olica de Valpara\'iso, Casilla 4059, Valpara\'iso, Chile.\\
${}^{5}$Institute of Gravitation and Cosmology, Department of Physics, School of Mathematics and Physics, Bohai University, Jinzhou 121013, China.\\
${}^{6}$State Key Laboratory of Theoretical Physics, Institute of Theoretical Physics, Chinese Academy of Sciences, Beijing 100190, China.}

\begin{abstract}
We investigate the holographic fermionic phase transition induced by the effective impurity in holography,
which is introduced by massless scalar fields in Einstein-Maxwell-massless scalar gravity.
We obtain a phase diagram in $(\alpha, T)$ plane separating the Fermi liquid phase and the non-Fermi liquid phase.
\end{abstract}

\maketitle
\section{Introduction}
Holographic duality (gauge/gravity correspondece) \cite{adscft,gkp,w} is a particular kind of strong/weak duality, which gives us a  powerful method to study strongly correlated systems in condensed matter physics via the study of the black hole physics with one extra dimension. Especially, it provides a novel way to explore the strongly correlated fermionic system.

The early works \cite{f1,f2,f3,f4,IL} on this subject have made efforts on understanding the mysterious behaviors of the existing non-Fermi liquid which is named as ``holographic non-Fermi liquid". In detail, by studying the evolution of the bulk Dirac equation in the RN-AdS black hole with the ingoing boundary condition at the horizon, the fermionic correlator can be extracted at the AdS boundary holographically. This proposal attracts many interests in the related topics.  Some endeavors have been made to generalize the investigations of the probe fermion to the general charged Gauss-Bonnet black hole, charged dilatonic black hole, charged Lifshitz black hole and so on\cite{JPW1,kuang1,kuang2,JPW2,FLQ1,FLQ2,FLQ3}. One also implemented the holographic non-relativistic fermionic fixed points by imposing the Lorentz violating boundary condition, which results in an infinite flat band in the dual boundary field theory\cite{1108.1381,1110.4559,1409.2945}. By introducing the interaction between Dirac field and gauge field, it was disclosed that the dual liquid can transform from the Fermi liquid to non-Fermi liquid and also to the Mott insulating phase\cite{1010.3238,1012.3751,1102.3908,1405.1041,1404.4010,1411.5627}.
The properties of the fermionic spectral function including lattice effects, which is introduced in \cite{1209.1098,1311.3292,1309.4580}, has been explored in \cite{1304.2128,1410.7323}.
It shows some crucial characteristics, including periodic structure and Brillouin zones,
which can be compared with the results of Angle-resolved photoemission spectroscopy (ARPES) or Scanning tunneling microscopy (STM), in real condense physical system.

In real condensed matter system, another crucial role is impurity.
The physical properties of materials will drastically change when impurity is added. A famous example is the cuprates, which exhibits Mott insulating phase, non-Fermi liquid phase and Fermi liquid phase as the doping levels change.
Another well-known example is the Kondo effect that when the magnetic impurity is introduced, the resistance increases logarithmically with the decrease of the temperature.
In holography, impurity can be introduced in Einstein-Maxwell-massless scalar gravity \cite{massless1,massless2},
in which it is implemented by spatially dependent sources for scalar field operators with being linear in coordinates\footnote{Impurity is also implemented by disorder in holography. For instance, one can refer to \cite{HartnollHS,ZengDRA}.}.
By introducing an extra charged complex scalar field, the authors of \cite{KimDNA,CLGBM} built a holographic superconductor with momentum relaxation.
They found a new type of superconductor induced by the effective impurity $\alpha$ even at vanishing chemical potential $\mu=0$.
More interestingly, the effective impurity $\alpha$ in holography can induce a coherent/incoherent metal transition \cite{massless1,massless3}.
The coherent metal has a standard Drude peak, which is usually described by Fermi liquid theory,
while the incoherent metal follows non-Drude behavior, which complies with non-Fermi liquid theory.
This motivates us to investigate the fermionic response with impurity in holography.

Subsequently, we will shortly review the simple holographic model of momentum relaxation \cite{massless1,massless2} and discuss its near horizon geometry at any temperature in section II.
Then the equation of motion and Green function on this background are derived.
Afterwards, we exhibit our numerical results and discuss Fermi surface structure and phase diagram induced by the effective impurity parameter $\alpha$.
Finally, our results are summarized in section V.

\section{Einstein-Maxwell gravity with massless scalar field sources}

We start with a brief review of the bulk Einstein-Maxwell gravity accompanied with  massless scalar field sources and analyze its near horizon geometry.

\subsection{Background geometry with scalar field sources}
We combine the action of free massless scalars together with  the  Einstein-Maxwell  action  in  $d+1$ dimension \cite{massless1},
\begin{equation}\label{EMaction}
S = \frac{1}{2\kappa^2}\int_M d^{d+1} x \sqrt{-g} \left[ R - 2 \Lambda  - \frac{1}{4} F_{\mu\nu}F^{\mu\nu} - \frac{1}{2} \sum_{I}^{d-1} (\partial \Psi_I)^2\right]- \frac{1}{2\kappa^2}\int_{\partial M}d^dx \sqrt{-\gamma} 2K,
\end{equation}
where $\Lambda = - d(d-1)/(2 L^2)$ and the field strength $F_{\mu\nu}=\partial_{\mu} A_{\nu}-\partial_{\nu} A_{\mu}$ for a $U(1)$ gauge field. The second term  is the Gibbons-Hawking term where $\gamma$ is the induced metric on the boundary and $K$ is the trace of the extrinsic curvature.

For simplicity, we choose the gravitational coupling $2\kappa^2=16\pi G_{d+1}$ and the AdS radius $L$ as unity.
The equations of motion can be easily obtained as
\begin{align}\label{actioneom}
R_{\mu\nu}-\frac{1}{2}g_{\mu\nu} \left( R - 2 \Lambda -\frac{1}{4}F_{\mu\nu}F^{\mu\nu}  - \frac{1}{2}\sum_{I=1}^{d-1} (\partial\Psi_I)^2 \right)
-\frac{1}{2}  \sum_{I=1}^{d-1}  \partial_{\mu} \Psi_I \partial_{\nu} \Psi_I -\frac{1}{2} {F_{\mu}}^{\rho} F_{\nu\rho}=0,\\
\nabla_{\mu} F^{\mu\nu} =0,~~~~~~\nabla^2 \Psi_I =0.
\end{align}
The equations of motions admit the following isotropic solution of various fields
\footnote{Note that it was addressed in \cite{massless1} that this solution is highly similar to the solution proposed in massive gravity\cite{1301.0537,1306.5792,1308.4970}.}
\begin{align}\label{ansatz}
ds^2 = - f(r) dt^2 + \frac{dr^2}{f(r)} + r^2 \sum_{i=1}^{d-1} dx^i dx^i,\nonumber\\
f = r^2  - \frac{\alpha^2}{2(d-2)} - \frac{m_0}{r^{d-2}} +\frac{Q^2}{r^{2(d-2)}},\nonumber\\
A = \mu \left(1 - \frac{r_0^{d-2}}{r^{d-2}}\right) dt,~~~~~\Psi_I = \alpha_{Ii} x^i=\alpha\delta_{Ii}x^{i},
\end{align}
where
\begin{align}\label{m0Qmu}
m_0 = r_0^d \left(1+ \frac{d-2}{2(d-1)}\frac{\mu^2}{r_0^2}- \frac{1}{2(d-2)}\frac{\alpha^2}{r_0^2} \right), \\
Q^2=\frac{(d-2)\mu^2r_0^{2(d-2)}}{2(d-1)},
\end{align}
and $\alpha^2 \equiv \frac{1}{d-1}\sum_{i= 1}^{d-1} \vec\alpha_i \cdot \vec \alpha_i$ with the vector notation $(\vec \alpha_i)_I = \alpha_{I i}$ and  $\vec\alpha_i \cdot \vec \alpha_j = \sum_I \alpha_{Ii} \alpha_{Ij}$\footnote{It is worthy pointing out that in order to consider the CFT dual to
the anisotropic system, we have to focus on the dimension $d\geqslant3$.  Specially,  in the background with dimension $d=1$,
 a logarithmic divergence will appear and $\alpha$ cannot be introduced into the action.}. We can see that due to the presence of the spatial dependence in the massless
scalar field, the full solution  is not isotropic and homogeneous, though the metric is isotropic. Note that the anisotropic solutions with only one scalar field has been studied in \cite{CL2,CL3}. The parameter $\alpha$ mimics the strength of the
translational symmetry, which means the momentum relaxation in the dual theory. Note that when $\alpha=0$, the solution goes
back to the RN-AdS black hole dual to the field theory with translational invariance.

The temperature and entropy density of the black hole are given by
\begin{align}
\label{T}
T &= \frac{f'(r_0)}{4 \pi} = \frac{d r_0}{4 \pi} \left(1- \frac{\alpha^2}{2dr_0^2}  -  \frac{(d-2)^2Q^2}{d r_0^{2d-2}}   \right),\\
\label{S}
s &= 4 \pi r_0^{d-1}.
\end{align}
The solution is fully characterized by $T$, $\mu$ and $\alpha$. For a given chemical potential and temperature, we can determine the value of $\alpha$.
This model can be seen as the case of the long wavelength limit of Q-lattice model \cite{1311.3292}.
But an periodic lattice structure is absent because the scalar source is linearly dependent on the spatial coordinates. Therefore, it is more reasonable to treat $\alpha$ as impurity, as illustrated in \cite{massless1,massless3}.

\subsection{Near horizon geometry}
In order to study the near horizon geometry at any temperature, we will introduce a length scale $r_{*}$ by parameterizing $Q$ with dimension $d-1$ as
\begin{align}\label{rstar}
Q=\sqrt{1- \frac{\alpha^2}{2dr_0^2}}\sqrt{\frac{d}{d-2}}r_{*}^{d-1}.
\end{align}
Then the temperature can be written as
\begin{align}
T = \frac{d r_0}{4 \pi}\left(1- \frac{\alpha^2}{2dr_0^2}\right)\left(1-(\frac{r_{*}}{r_0})^{2d-2}\right).
\end{align}
We can see that $r_{*}$ is equal to the horizon radius $r_{0}$ at zero temperature and smaller than the radius $r_{0}$ at finite temperature.
To work out the near horizon geometry, we consider the scaling limits\cite{f1}
\begin{align}\label{scalinglimit}
r-r_{*}=\omega\frac{L_{2}^{2}}{\zeta},~~~,r_{0}-r_{*}=\omega\frac{L_{2}^{2}}{\zeta_0},~~~t=\omega^{-1}\tau
\end{align}
with $\omega\rightarrow 0$, $\zeta$ and $\tau$ finite. $L_2$  is the curvature radius of $AdS_2$ geometry whose expression will be shown later.

Under the above transformation, the leading term of the redshift factor $f$ in \eqref{ansatz}  is
\begin{align}
f&=\omega^2L_{2}^2\left(\frac{1}{\zeta_0^2}-\frac{1}{\zeta^2}\right)
\end{align}
with the curvature radius $L_{2}^2 =\frac{1}{d(d-1)} \frac{(d-1)\alpha^2 + (d-2)^2\mu^2}{\alpha^2 + (d-2)^2\mu^2}$.  Thus, metric near horizon becomes
\begin{equation}\label{ads2ft}
ds^2=\frac{L_{2}^{2}}{\zeta^2}\left[-\left(1-\frac{\zeta^2}{\zeta_0^2}\right)d\tau^2+\frac{1}{\left(1-\frac{\zeta^2}{\zeta_0^2}\right)}d\zeta^2\right]+r_{*}^{2}\sum_{i=1}^{d-1} dx^i dx^i.
\end{equation}
and the temperature is  $T=\frac{1}{2\pi\zeta_0}$.
The zero temperature limit will be reached as $\zeta_0\rightarrow\infty$ so that the formulation of metric can be deduced from Eq. \eqref{ads2ft} as
\begin{equation}
ds^2=\frac{L_{2}^{2}}{\zeta^2}\left(-d\tau^2+d\zeta^2\right)+r_{*}^{2}\sum_{i=1}^{d-1} dx^i dx^i.
\end{equation}
which is the well known $AdS_2\times \mathbb{R}^{d-1}$ metric at zero temperature.

\section{Holographic Fermionic implementation}

\subsection{The flow equation}
In order to study the  Fermionic theory dual to this bulk theory, we consider a probe spinor field with the charge $q$ and mass $m$, which is dual to fermionic operator $\mathcal{O}$ in boundary $CFT_{d}$ with conformal dimension $\Delta=m + \frac{d}{2}$. The action of spinor field is given by
\begin{equation}\label{bulkaction}
S_{F}=\int
d^{d+1}x\sqrt{-g}i\bar{\psi}\Big(\Gamma^a D_{a}-m \Big)\psi,
\end{equation}
where $\Gamma^a=(e_\mu)^a\Gamma^\mu$ and the covariant derivative $D_{a}=\partial_{a}+\frac{1}{4}(\omega_{\mu\nu})_a\Gamma^{\mu\nu}-iq_{i}A_{a}$
with  $\Gamma^{\mu\nu}=\frac{1}{2}[\Gamma^\mu,\Gamma^\nu]$, the spin connection $(\omega_{\mu\nu})_{a}=(e_\mu)^b\nabla_a(e_\nu)_b$, where
$(e_\mu)^{a}$ forms a set of orthogonal normal vector bases. From the above Eq. \eqref{bulkaction}, the equation of motion of Dirac field can be written as
\begin{equation}\label{Diraceom1}
\Gamma^a D_{a}\psi-m\psi=0.
\end{equation}
In order to investigate the Dirac equation in the Fourier space, we use the translational invariance $\psi=\int d\omega dk e^{-i\omega t+ik_{i}x^{i}}\psi$. Then, we can rescale the Dirac field by $\psi=(-g g^{rr})^{-\frac{1}{4}}\phi$, and  remove the spin connection completely in the Dirac equation.
Considering the rotation symmetry in the spatial directions, we can simply set $k_i=k_x$. Then, it is convenient to consider the following basis
\begin{eqnarray}\label{GammaMatrices}
 && \Gamma^{r} = \left( \begin{array}{cc}
-\sigma^3 \textbf{1} & 0  \\
0 & -\sigma^3 \textbf{1}
\end{array} \right), \;\;
 \Gamma^{t} = \left( \begin{array}{cc}
 i \sigma^1 \textbf{1}  & 0  \\
0 & i \sigma^1 \textbf{1}
\end{array} \right),  \;\;
\Gamma^{x} = \left( \begin{array}{cc}
-\sigma^2 \textbf{1}  & 0  \\
0 & \sigma^2 \textbf{1}
\end{array} \right),
\phi=\left(\begin{array}{c}\phi_1 \\ \phi_2\\ \end{array} \right),
\end{eqnarray}
where $\phi_I$ are two-component spinors, $\textbf{1}$ is an identity matrix of size $2^{\frac{d-3}{2}}$ for odd $d$ (or $2^{\frac{d-4}{2}}$ for even $d$).
By taking the above operations, the Dirac equation (\ref{Diraceom1}) can be written as
\begin{eqnarray}\label{phi12eom}
\frac{\sqrt{g_{xx}}}{\sqrt{g_{rr}}}\partial_r
\left(\begin{array}{c}\phi_1 \\ \phi_2\\ \end{array} \right)
+\sqrt{g_{xx}}m\sigma^{3}\otimes
\left(\begin{array}{c}\phi_1 \\ \phi_2\\ \end{array} \right)=\frac{\sqrt{g_{xx}}}{\sqrt{g_{tt}}}(\omega+q A_{t})i\sigma^{2}\otimes
\left(\begin{array}{c}\phi_1 \\ \phi_2\\ \end{array} \right)
\mp k\sigma^{1}\otimes
\left(\begin{array}{c}\phi_1 \\ \phi_2\\ \end{array} \right)
\end{eqnarray}
Furthermore, by setting $\phi_I=\left(\begin{array}{c}y_I \\ z_I\\\end{array} \right)$ and introducing the ratios $\xi_I=\frac{y_I}{z_I}$,
the equation of motion can be decoupled and reduced to
\begin{equation}\label{floweqa}
(\frac{\sqrt{g_{xx}}}{\sqrt{g_{rr}}}\partial_r
+2\sqrt{g_{xx}}m)\xi_{I}
=[\frac{\sqrt{g_{xx}}}{\sqrt{g_{tt}}}(\omega+q A_{t})+(-1)^{I}k]+[\frac{\sqrt{g_{xx}}}{\sqrt{g_{tt}}}(\omega+q A_{t})-(-1)^{I}k]\xi_{I}^{2},
\end{equation}
The in-falling boundary condition at the horizon for $\omega\neq0$ is
\begin{equation}\label{bdycond}
\xi_I\buildrel{r \to r_0}\over =i.
\end{equation}

\subsection{Near boundary and boundary conditions}
Near boundary $r\rightarrow\infty$, the metric has the behavior as
\begin{equation}\label{metricbdy}
g_{tt}\rightarrow r^2,~~g_{rr}\rightarrow \frac{1}{r^2},~~g_{ii}\rightarrow r^2.
\end{equation}
Substituting the near boundary metric into Eq. \eqref{phi12eom}, we have
\begin{equation}\label{Diraceqbdy}
\left[r^2\partial_r+r m\sigma^3-(\omega+q\mu)i\sigma^2-(-1)^Ik\sigma^1 \right]\phi_I =0.
\end{equation}
Therefore, the Dirac equation becomes
\begin{equation}
\left(r^2\partial_r+r m\sigma^3 \right)\phi_I =0.
\end{equation}
The solution of the Dirac equation near the AdS boundary is
\begin{equation}\label{boundarycondition}
\phi_I\xrightarrow[ ]{r\rightarrow\infty} a_I r^{-m}\left(
\begin{array}{c}
1 \\0 \\ \end{array}
 \right)+b_Ir^{m}\left(
\begin{array}{c} 0 \\ 1 \\\end{array}\right),
\end{equation}
where $a_I$ and $b_I$ can be regarded as the  response and the source, respectively. Then, we suppose the source and the response in Eq. \eqref{boundarycondition} are related by
\begin{equation}
a_I \left(
\begin{array}{c}
 1 \\0 \\\end{array}
 \right)= \mathcal{S}b_I
 \left(\begin{array}{c} 0 \\1 \\
\end{array}\right),
\end{equation}
the boundary Green¡¯s functions $G(\omega,k)$ is given by \cite{IL}
\begin{equation}
G=-i\mathcal{S}\gamma^0.
\end{equation}
Recalling  the definition of $\xi_I=\frac{y_I}{z_I}$, we can write the Green function  as
\begin{equation}\label{gi}
G(\omega,k)=\lim_{r\rightarrow\infty}r^{2m}
    \left(
    \begin{array}{cc}
    \xi_1 & 0 \\
     0 & \xi_2 \\
    \end{array}
    \right),
\end{equation}
and the behavior of $\xi_I$ near the boundary is
\begin{equation}
\xi_{I}\buildrel{r \to \infty}\over =r^{-2m}G_{II}.
\end{equation}

\subsection{Spinor field at the near-horizon geometry and the dispersion relation}
Near the horizon region, the geometry approaches $AdS_{2}\times \mathbb{R}^{d-1}$  which is controlled by Eq. \eqref{ads2ft}.
Substituting Eq. (\ref{ads2ft}) into Eq. (\ref{phi12eom}),  the Dirac equation Eq. \eqref{phi12eom}  can be expressed as
\begin{eqnarray}
\partial_{\zeta}\phi_{I}(\zeta)=\frac{L_{2}}{\zeta}\left(\frac{\zeta_0}{\zeta_0-\zeta}\right) m \sigma^{3}\phi_{I}(\zeta)
-i\Big[\left(\frac{\zeta_0}{\zeta_0-\zeta}\right)^2+\frac{\mu q (d-2)L_{2}^{2}}{r_*\zeta}\left(\frac{\zeta_0}{\zeta_0-\zeta}\right)\Big]\sigma^{2}\phi_{I}(\zeta)\nonumber\\
-(-1)^{I}\frac{k}{r_{*}}\frac{L_{2}}{\zeta}\left(\frac{\zeta_0}{\zeta_0-\zeta}\right)\sigma^{1}\phi_{I}(\zeta).
\end{eqnarray}
Near the $AdS_{2}$ boundary $\zeta\rightarrow 0$, the above equation can be simplified into
\begin{eqnarray}\label{phi0}
\zeta \partial_{\zeta}\phi_{I}(\zeta)=-U\phi_{I}(\zeta),~~
U=\left(
    \begin{array}{cc}
    m L_2 & \tilde{m}_{I}L_2- \frac{\mu q (d-2)L_{2}^{2}}{r_*} \\
     \tilde{m}_{I}L_2+ \frac{\mu q (d-2)L_{2}^{2}}{r_*} & -m L_2 \\
    \end{array}
    \right),
\end{eqnarray}
with $\tilde{m}_{I}=-(-1)^{I}\frac{k}{r_{*}}$.
Following the discussion in \cite{f1}, equation (\ref{phi0})
is nothing but the equation of motion for spinor
fields in the $AdS_2$ background with masses $[m,\tilde{m}_{I}]$ with $\tilde{m}_{I}(I=1,2)$ the time-reversal violating mass
terms. In this case, the spinor field is dual  to a spinor
operators in the IR $CFT_{1}$ whose conformal
dimensions of the operator are $\delta=\nu_{k}+\frac{1}{2}$ with
\begin{equation}\label{nuk}
\nu_{k}=\sqrt{m^{2}L_{2}^{2}+\tilde{m}_{I}^{2}L_{2}^{2}-\Big(\frac{\mu q(d-2)L_{2}^{2}}{r_{*}}\Big)^{2}}.
\end{equation}

For the real $\nu_k$, the scaling exponent $z$ of dispersion relation of our UV Green function $\rm {Im}G_{22}$ can be determined by\cite{f3}
\begin{equation}\label{scalexp}
z= \left\{ \begin{array}{ll}
1 & \textrm{~~~with $\nu_{k_{F}}>\frac{1}{2}$}\\
\frac{1}{2\nu_{k_{F}}} & \textrm{~~~with $\nu_{k_{F}}<\frac{1}{2}$}
\end{array} \right.
\end{equation}
where $\nu_{k_{F}}$ can be obtained from Eq. (\ref{nuk}) at $k=k_F$. The scaling exponent $z$ contains important property of the Green
function near the Fermi momentum $k_{\perp}=k-k_{F}$ and it describes the excitation of dispersion relation
\begin{equation}\label{dr}
\omega_{\ast}(k_{\perp})\sim k_{\perp}^{z},
\end{equation}
where $\omega_{\ast}(k_{\perp})$ is the real part of  the complex frequency at which the Green function has a pole\cite{f2}.
After determining the Fermi momentum from numerical calculation, we can analytically compute the scaling exponent z of the dispersion relation through Eq. \eqref{nuk} and Eq. \eqref{scalexp}.
Note that $z=1$ is an necessary character for the Fermi liquid. So in the following discussion, $z$ moves from $1$ to other values
should be the distinguishing feature for the phase transition from Fermi liquid to non-Fermi liquid.
\section{Fermi surface and phase diagram}
In this section, we shall study the Fermi surface structure and the phase diagram ($\alpha$,T).
We will demonstrate that the transition from the Fermi liquid to the non-Fermi liquid happens when the impurity parameter $\alpha$ becomes larger.
In numerical calculation, we only focus on the Green's function $G_{22}(\omega,k)$ due to the symmetry $G_{22}(\omega,k)=G_{11}(\omega,-k)$,
which can be easily analyzed from Eq. (\ref{floweqa}) and the boundary condition Eq. (\ref{bdycond}). We set $m=0$, $q=1$ in most of our study unless we emphasize different setting. In addition, in the following discussion, we can set $r_0=1$ by rescaling.
\subsection{Results at zero temperature}
\begin{figure}
\centering
\includegraphics[width=.3\textwidth]{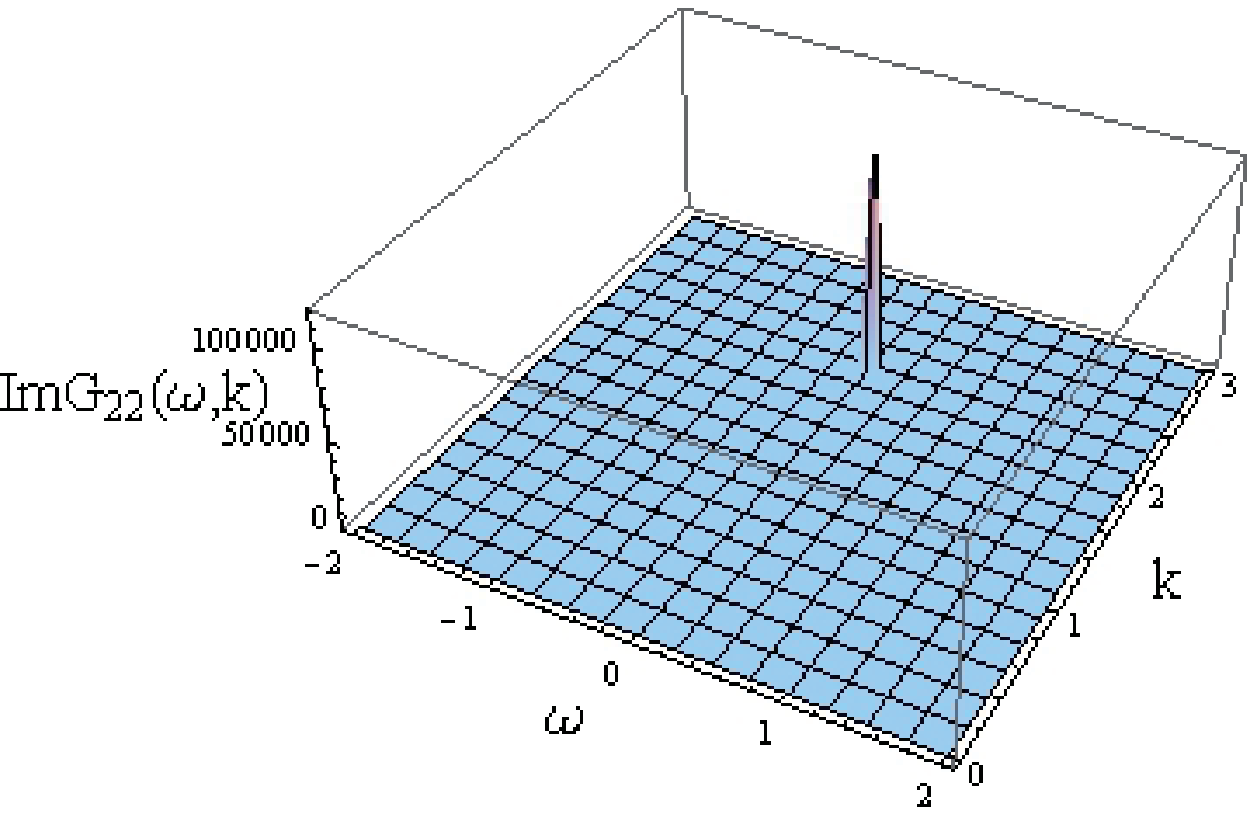}\hspace{0.5cm}
\includegraphics[width=.3\textwidth]{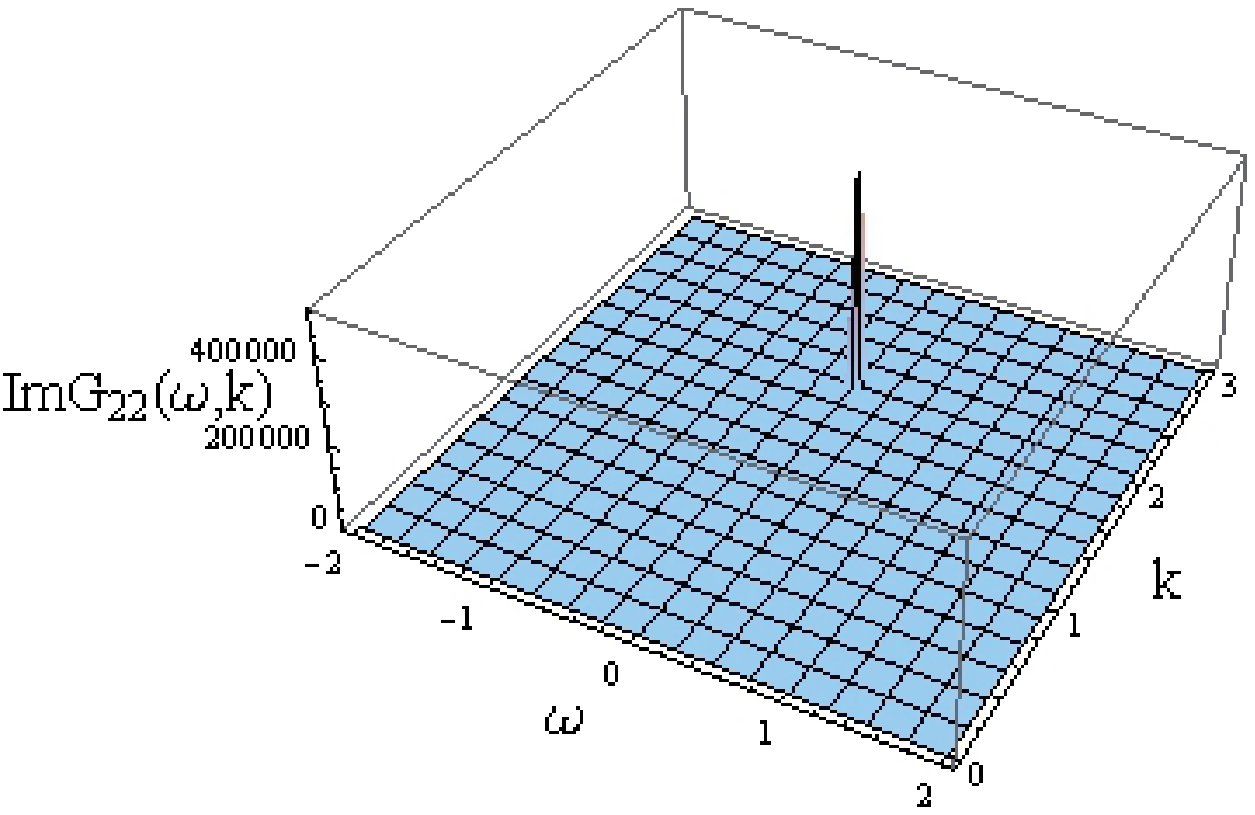}\hspace{0.5cm}
\includegraphics[width=.3\textwidth]{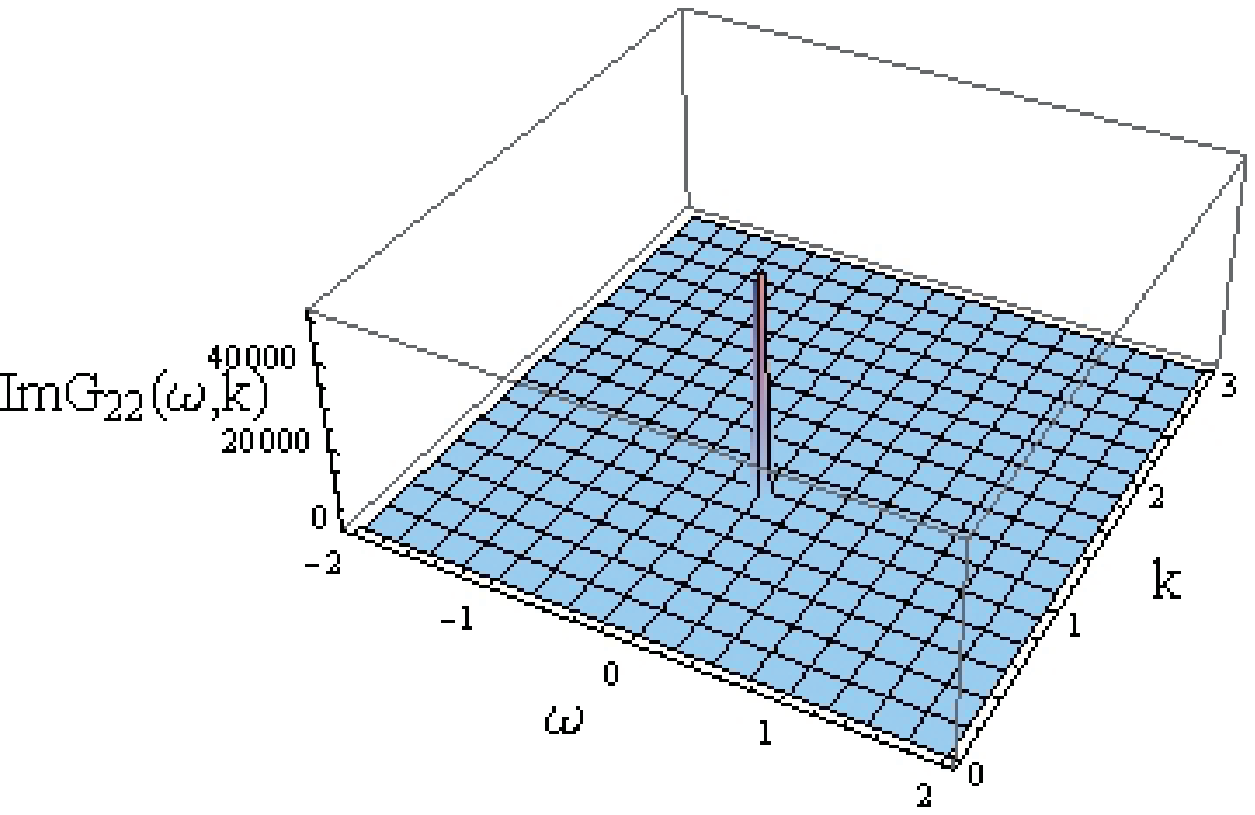}
 \caption{The 3D plot of Green function with $\alpha=0$(left), $\alpha=1$(middle) and $\alpha=2$(right), respectively. }
 \label{a3d}
\end{figure}

\begin{figure}
\centering
\includegraphics[width=.4\textwidth]{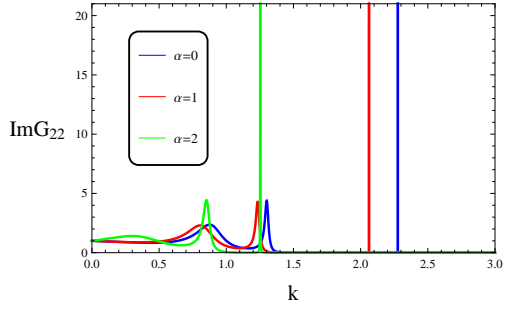}
 \caption{Green function at tiny frequency with different $\alpha$. }
 \label{kfaT0}
\end{figure}

In this subsection, we will study the Fermi surface structure  at zero temperature. We discuss the case with $d=3$ as the first step. From Fig.~\ref{a3d}, we can see that a sharp quasi particle like peak emerges near $\omega=0$ for different $\alpha$, which denotes the Fermi surface.
By definition, we can determine the Fermi momentum $k_F$ by locating the position of the maximal value of $ImG_{22}$ near $\omega=0$ as shown in Fig.~\ref{kfaT0}. We list some Fermi momentum $k_F$ with different $\alpha$ in Table \ref{akfnukz}.
We can see that with the increase of $\alpha$, the Fermi momentum $k_F$ becomes smaller.
After determining numerically the Fermi momentum $k_F$,
the scaling exponent $z$ of the dispersion relation can be calculated by Eq. (\ref{scalexp}) analytically.
The results have also been summarized in Table \ref{akfnukz}.
\begin{table}
\centering
\footnotesize
\begin{tabular}{|c|c|c|c|c|c|c|c|c|c|}
\hline
$d+1$ & \multicolumn{3}{c|}{4} & \multicolumn{3}{c|}{5} & \multicolumn{3}{c|}{6} \\
\cline{1-10}
$\alpha$ & 0& 1 & 2& 0& 1 & 2& 0& 1 & 2  \\
\hline
$k_F$ & 2.27694628& 2.06223335 & 1.2532708 & 1.8880638 & 1.7800883 & 1.3981258 & 1.7228794 & 1.6551679 & 1.4260239  \\
\hline
$\nu_{k}$ & 0.728524& 0.665326 & 0.377719 & 0.361108 & 0.338395 & 0.238421 & 0.220036 & 0.207121 & 0.151011   \\
\hline
$z$ & 1 & 1 & 1.32373 & 1.38463 & 1.47756 & 2.09713 & 2.27236 & 2.41405 & 3.31103   \\
\hline
\end{tabular}
\caption{$k_F$, $\nu_{k}$ and $z$ with different dimension and $\alpha$ ($T=0$).}
\label{akfnukz}
\end{table}

The effect of the dimension $d$ is explicitly reflected in the flow equation \eqref{nuk}. With careful numerical calculation, we list
the results with different dimensions in Table \ref{akfnukz}. For the same $\alpha$, Fermi momentum is suppressed in higher dimensional dual theory.
In addition, the scaling exponent $z$ of the dispersion relation decreases with the decrease of the spacetime dimension, which agrees well with the observation in Charged Gauss-Bonnet gravity in \cite{kuang1}.

From Table \ref{akfnukz}, an interesting result is that the transition from Fermi liquid phase to the non-Fermi liquid phase happens when the impurity $\alpha$ is beyond some critical value. This motivates us to turn on the temperature and draw the phase transition diagram of $\alpha-T$.

\subsection{Results at finite temperature}
\begin{figure}
\centering
\includegraphics[width=.4\textwidth]{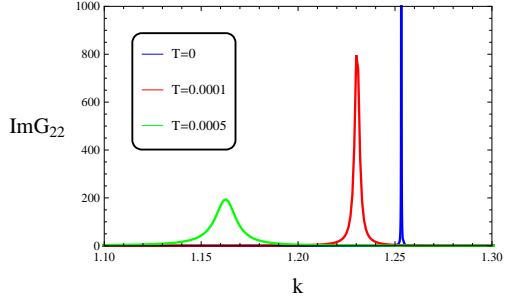}
 \caption{Green function at different temperatures with $\alpha=2$. }
 \label{kfTa2}
\end{figure}

\begin{figure}
\centering
\includegraphics[width=.4\textwidth]{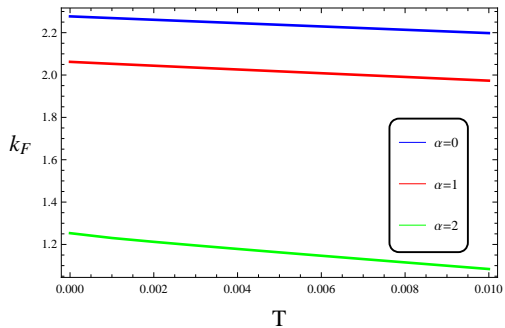}
 \caption{Fermi momentum versus temperature for $\alpha=0$(Blue), $\alpha=1$(Red) and $\alpha=2$(Green), respectively. }
 \label{kfaT}
\end{figure}

Now, we turn to explore the Fermi surface structure at the finite temperature and plot the phase diagram ($\alpha$, T).
Due to the thermal fluctuations at finite temperature, the height of the quasi-particle
peak becomes much lower and the width much broader (Fig.~\ref{kfTa2}).
Using the same numerical method as the case at zero temperature,
we present the Fermi momentum $k_F$ as the function of $T$ with different $\alpha$ in Fig.~\ref{kfaT}.
We can see that with the increase of the temperature, the Fermi momentum $k_F$ decreases for all $\alpha$.

\begin{figure}
\centering
\includegraphics[width=.4\textwidth]{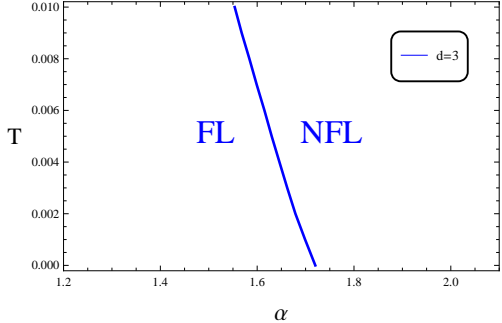}
 \caption{Phase diagram with $d=3$.}
 \label{Tad}
\end{figure}

Furthermore, we show the phase diagram $(\alpha,T)$ for $d=3$ in Fig.~\ref{Tad}. In Fig.~\ref{Tad},
the line denotes the marginal Fermi liquid phase and the left region is the Fermi liquid while the right region corresponds to the non-Fermi liquid.
It exhibits a phase transition from Fermi liquid to non-Fermi liquid induced by the effective impurity $\alpha$,
which is qualitatively consistent with the picture that a coherent/incoherent metal transition induced by $\alpha$ in \cite{massless1}.
We would like to point out that the phase diagram $(\alpha,T)$ exhibiting in Fig.~\ref{Tad} seems to be opposite to the one relating to the doping drawn in the high-$T_c$ cuprate superconductors\cite{htcsc}. This phenomenon somehow supports that
 the impurity $\alpha$ introduced here may not be considered as doping which is claimed in \cite{1504.05561}
 \footnote{The authors of \cite{1504.05561} treated  $\alpha$  in a more general model \cite{1411.1003}
 as disorder-strength but the existence of Anderson localization needs to  be further carefully confirmed.}.
At least,  $\alpha$ provides a new mechanism introducing impurity from holography.
It would certainly be interesting if such a phase diagram can be observed in some real materials in the future.

\begin{figure}
\centering
\includegraphics[width=.35\textwidth]{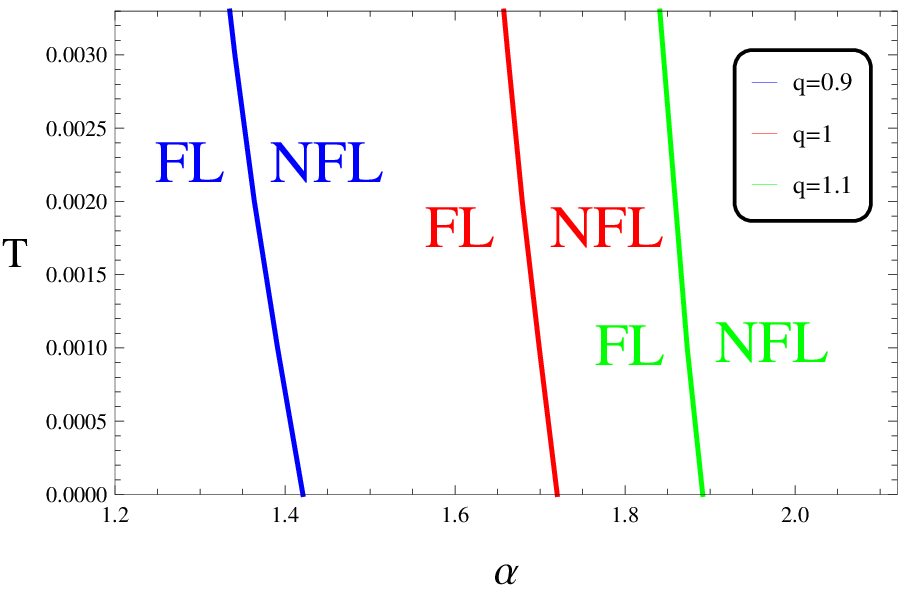}\hspace{0.5cm}
\includegraphics[width=.35\textwidth]{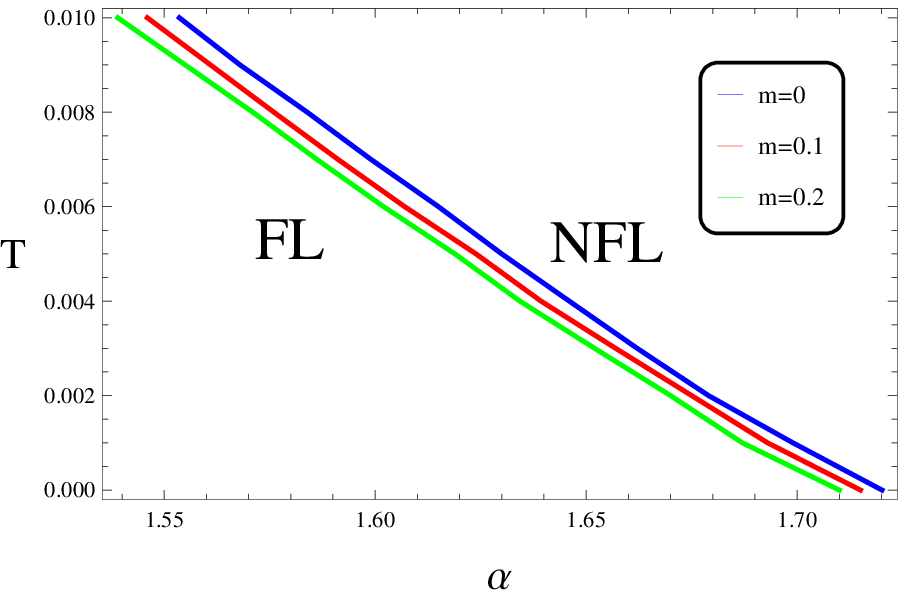}
 \caption{Left: Phase diagram for $d=3$ and $m=0$ with different fermion charge. The line denotes Marginal Fermi Liquid(MFL) where $\nu_k=\frac{1}{2}$ for $q=0.9$(Blue), $q=1$(Red), $q=1.1$(Green). Right: Phase diagram for $d=3$ and $q=1$ with different fermion mass . The line denotes MFL  for $m=0$(Blue), $m=0.01$(Red), $m=0.02$(Green).
 In each case, the left region of each line denotes Fermi liquid while the right side means non-Fermi liquid.}
 \label{a-Tqm}
\end{figure}

As is known that the model parameters $m$ and $q$ have effects on the types of dual liquids and the phase transition from Fermi to non-Fermi liquids in the holographic fermionic systems \cite{f3,f4,IL,1409.2945}. It was addressed that with fermion charge $q$ increasing, the Fermi momentum increases approximately linearly and the scaling exponent $z$ of dispersion relation decreases rapidly. Thus, the dual liquid changes from the Landau Fermi liquid type to the non-Fermi liquid is more difficult for larger charge. Here, we find the similar tendency in our model. In the left plot of Fig.~\ref{a-Tqm}, we see that in the case with larger charge, the phase transition happens with stronger impurity parameter $\alpha$. The reason is why larger charge is harder to make the phase transition happen so that $q$ suppresses the effect of impurity. We move on to discuss the effect of the mass of fermions. For a given $\alpha$, the Fermi momentum $k_F$ decreases and the scaling exponent increase with the increase of $m$ which agrees well with the results in \cite{1409.2945}. At certain $\alpha$, we always find a phase transition from the Fermi liquid to the non-Fermi liquid with fixed mass. From the right plot of Fig.~\ref{a-Tqm}, we see that at a fixed temperature, the critical $\alpha$ is smaller for larger mass, namely that larger mass promotes the phase transition.
We also study the phase transition in higher dimension which may affect the type of the dual liquid.
An interesting result is that the Fermi liquid phase is forbidden for $\alpha^2>0$ in the $d=4$ and $d=5$ geometry.
And it can be realized when we choose a negative enough $\alpha^2$, which enhances the chemical
potential of the system with a fixed temperature.  However, this means that $\alpha$ should be imaginary,
which should be further understood from holography.

\section{Discussion and Conclusions}
In this paper, we have explored the Fermi surface and the phase diagram $(\alpha, T)$ in the holographic framework
based on the free probe fermions in Einstein-Maxwell gravity with spacial dependent massless scalar field, which introduces a new impurity effects from holography in the dual field theory.
The Fermi momentum was suppressed by the larger impurity parameter $\alpha$
while the conformal dimension of the Dirac field decreases as the impurity $\alpha$ increases at any temperature.
Furthermore, we have demonstrated that the effective impurity parameter $\alpha$ can induce the phase transition from the Fermi liquid to the Non-Fermi liquid.
This phenomenon results from that  large positive $\alpha$ corresponds to lower chemical potential in the dual system with fixed temperature.

Our work is a first step in studying the impurity effects on the fermionic spectral function in holography,
in which the influence of the effective impurity or the spatial broken translation symmetry only reflects on the Dirac equation through the metric correction.
To gain more insights into the impurity effects of the fermionic spectral function in holography,
we can further introduce an interaction between the massless scalar field and the Dirac field.
This can produce rich Fermi surface structure,
which can be compared with the results of Angle-resolved photoemission spectroscopy (ARPES) or Scanning tunneling microscopy (STM) in some real impurity materials,
and more complete phase diagram including Mott insulating phase.
The related work is under progress.
It is also interesting to consider the non-relativistic Fermionic fixed point to see the effect of the impurity $\alpha$ on the flat band.
We will present our results on this topic elsewhere in the future.

\begin{acknowledgments}
This work was supported by Natural Science Foundation of China. We would like to thank Matteo Baggioli for helpful suggestions and comments. L. Q. Fang is grateful to Xian-Hui Ge for  supports and discussions. L. Q. Fang is partly supported by the China postdoctoral Science Foundation under
Grant No. 2015M571551.  X. M. Kuang is partly supported by ARISTEIA II action of the operational programme education
and long life learning which is co-funded by the European Union (European Social Fund) and National Resources. X. M. Kuang is also supported by the Chilean FONDECYT Grant No.~3150006.
J. P. Wu is supported by the Natural Science Foundation of China under Grant
Nos. 11305018 and 11275208 and also supported by Program for Liaoning Excellent Talents in University (No. LJQ2014123).
\end{acknowledgments}


\begin{thebibliography}{99}
\bibitem{adscft}
J. M. Maldacena, The Large N Limit of Superconformal Field Theories and Supergravity, Adv. Theor. Math. Phys.  2 (1998) 231,
{[arXiv:hep-th/9711200]}.

\bibitem{gkp}
S. S. Gubser, I.R. Klebanov and A.M. Polyakov, Gauge Theory Correlators from Non-Critical String Theory, Phys. Lett.
B 428 (1998) 105, {[arXiv:hep-th/9802109]}.

\bibitem{w}
E. Witten, Anti De Sitter Space And Holography, Adv. Theor. Math. Phys. (1998) 253,
{[arXiv:hep-th/9802150]}.

\bibitem{f1}S. S. Lee, A Non-Fermi Liquid from a Charged Black Hole; A Critical Fermi Ball,
Phys. Rev. D 79 (2009) 086006,
[arXiv:0809.3402 [hep-th]].
\bibitem{f2} H. Liu, J. McGreevy and D. Vegh, Non-Fermi liquids from holography,
 Phys. Rev. D 83 (2011) 065029,
[arXiv:0903.2477 [hep-th]].
\bibitem{f3} T. Faulkner, H. Liu, J. McGreevy and D. Vegh, Emergent quantum criticality, Fermi surfaces, and AdS2, Phys. Rev. D 83 (2011) 125002,
[arXiv:0907.2694 [hep-th]].
\bibitem{f4} M. Cubrovic, J. Zaanen and K. Schalm, String Theory, Quantum Phase Transitions and the Emergent Fermi-Liquid, Science 325 (2009) 439,
[arXiv:0904.1993 [hep-th]].
\bibitem{IL}N. Iqbal and H. Liu, Real-time response in AdS/CFT with application to spinors,
Fortsch. Phys. 57, 367 (2009),
[arXiv:0903.2596 [hep-th]].
\bibitem{JPW1}J. P. Wu,  Holographic fermions in charged Gauss-Bonnet black hole,  JHEP 07 (2011) 106,
[arXiv:1103.3982 [hep-th]].
\bibitem{kuang1}X. M. Kuang, B. Wang, J. P. Wu, Dipole coupling effect of holographic fermion in the background of charged Gauss-Bonnet AdS black hole,
JHEP 1207(2012)125, [arXiv:1205.6674[hep-th]].
\bibitem{kuang2}
X. M. Kuang, B. Wang, J. P. Wu, Dynamical gap from holography in the charged dilaton black hole,Class. Quantum Grav. 30 (2013) 145011.
\bibitem{JPW2}J. P. Wu, Some properties of the holographic fermions in an extremal charged dilatonic black hole
, Phys. Rev. D 84,064008(2011),
[arXiv:1108.6134 [hep-th]].

\bibitem{FLQ1}L. Q. Fang, X. H. Ge, X. M. Kuang,
Holographic fermions in charged Lifshitz theory, Phys. Rev. D {\bf 86}, 105037 (2012).

\bibitem{FLQ2} L. Q. Fang, X. H. Ge, X. M. Kuang,
Holographic fermions with running chemical potential and dipole coupling, Nucl. Phys. B 877, 807 (2013).

\bibitem{FLQ3} L. Q. Fang, X. H. Ge, J. P. Wu, H. Q. Leng,
Anisotropic Fermi surface from holography, Phys. Rev. D {\bf 91}, 126009 (2015).

\bibitem{1108.1381}
  J.~N.~Laia and D.~Tong,
  A Holographic Flat Band,
  JHEP {\bf 1111} (2011) 125
  [arXiv:1108.1381 [hep-th]].

\bibitem{1110.4559}
  W.~J.~Li and H.~b.~Zhang,
  Holographic non-relativistic fermionic fixed point and bulk dipole coupling,
  JHEP {\bf 1111} (2011) 018
  [arXiv:1110.4559 [hep-th]].
\bibitem{1409.2945}
  X.~M.~Kuang, E.~Papantonopoulos, B.~Wang and J.~P.~Wu,
  Formation of Fermi surfaces and the appearance of liquid phases in holographic theories with hyperscaling violation,
  JHEP {\bf 1411} (2014) 086
  [arXiv:1409.2945 [hep-th]].


\bibitem{1010.3238}
  M.~Edalati, R.~G.~Leigh and P.~W.~Phillips,
  Dynamically Generated Mott Gap from Holography,
  Phys.\ Rev.\ Lett.\  {\bf 106} (2011) 091602
  [arXiv:1010.3238 [hep-th]].

\bibitem{1012.3751}
  M.~Edalati, R.~G.~Leigh, K.~W.~Lo and P.~W.~Phillips,
  Dynamical Gap and Cuprate-like Physics from Holography,
  Phys.\ Rev.\ D {\bf 83} (2011) 046012
  [arXiv:1012.3751 [hep-th]].

 \bibitem{1102.3908}
  D.~Guarrera and J.~McGreevy,
  Holographic Fermi surfaces and bulk dipole couplings,
  arXiv:1102.3908 [hep-th].

\bibitem{1411.5627}
  X.~M.~Kuang, E.~Papantonopoulos, B.~Wang and J.~P.~Wu,
  Dynamically generated gap from holography in the charged black brane with hyperscaling violation,
  JHEP {\bf 1504} (2015) 137
  [arXiv:1411.5627 [hep-th]].

\bibitem{1405.1041}
G. Vanacore, P. W. Phillips, Minding the Gap in Holographic Models of Interacting Fermions, Phys. Rev. D {\bf 90}, 044022 (2014), [arXiv:1405.1041].

\bibitem{1404.4010}
J. Alsup, E. Papantonopoulos, G. Siopsis, K. Yeter, Duality between zeroes and poles in holographic systems with massless fermions and a dipole coupling,
Phys. Rev. D {\bf 90}, 126013 (2014),[arXiv:1404.4010].


\bibitem{1209.1098}
  G.~T.~Horowitz, J.~E.~Santos and D.~Tong,
  Further Evidence for Lattice-Induced Scaling,
  JHEP {\bf 1211} (2012) 102
  [arXiv:1209.1098 [hep-th]].

\bibitem{1311.3292}
  A.~Donos and J.~P.~Gauntlett,
  Holographic Q-lattices,
  JHEP {\bf 1404} (2014) 040
  [arXiv:1311.3292 [hep-th]].

\bibitem{1309.4580}
  Y.~Ling, C.~Niu, J.~P.~Wu and Z.~Y.~Xian,
  Holographic Lattice in Einstein-Maxwell-Dilaton Gravity,
  JHEP {\bf 1311} (2013) 006
  [arXiv:1309.4580 [hep-th]].

\bibitem{1304.2128}
  Y.~Ling, C.~Niu, J.~P.~Wu, Z.~Y.~Xian and H.~b.~Zhang,
  Holographic Fermionic Liquid with Lattices,
  JHEP {\bf 1307} (2013) 045
  [arXiv:1304.2128 [hep-th]].

\bibitem{1410.7323}
  Y.~Ling, P.~Liu, C.~Niu, J.~P.~Wu and Z.~Y.~Xian,
  Holographic fermionic system with dipole coupling on Q-lattice,
  JHEP {\bf 1412} (2014) 149
  [arXiv:1410.7323 [hep-th]].



\bibitem{massless2}T. Andrade and B. Withers, A simple holographic model of momentum relaxation, JHEP 1405 (2014) 101,
[arXiv:1311.5157[hep-th]].
\bibitem{massless1} K. Y. Kim, K. K. Kim, Y. Seo and S.J. Sin, Coherent/incoherent metal transition in a holographic model, JHEP 1412 (2014) 170
[arXiv:1409.8346[hep-th]].

\bibitem{HartnollHS}
  S.~A.~Hartnoll and C.~P.~Herzog,
Impure AdS/CFT correspondence,
Phys.\ Rev.\ D {\bf 77}, 106009 (2008).
[arXiv:0801.1693 [hep-th]].

\bibitem{ZengDRA}
  H.~B.~Zeng and H.~Q.~Zhang,
  Zeroth Order Phase Transition in a Holographic Superconductor with Single Impurity,
Nucl.\ Phys.\ B {\bf 897}, 276 (2015).
[arXiv:1411.3955 [hep-th]].
\bibitem{KimDNA}
  K.~Y.~Kim, K.~K.~Kim and M.~Park,
  A Simple Holographic Superconductor with Momentum Relaxation,
JHEP {\bf 1504}, 152 (2015).
[arXiv:1501.00446 [hep-th]].
\bibitem{CLGBM}
  L.~Cheng,~X.~H.~Ge and Z.~Y.~Sun,
  Thermoelectric DC conductivities with momentum dissipation from higher derivative gravity,
  JHEP 04 (2015) 135.
  [arXiv:1411.5452]

\bibitem{massless3} B. Gout$\acute{\textmd{e}}$raux, Charge transport in holography with momentum dissipation, JHEP 1404
(2014) 181, [arXiv:1401.5436].
 \bibitem{1301.0537}
  D.~Vegh,
  Holography without translational symmetry,
  arXiv:1301.0537 [hep-th].
\bibitem{1306.5792}
  R.~A.~Davison,
  Momentum relaxation in holographic massive gravity,
  Phys.\ Rev.\ D {\bf 88} (2013) 086003,
  [arXiv:1306.5792 [hep-th]].
\bibitem{1308.4970}
  M.~Blake and D.~Tong,
  Universal Resistivity from Holographic Massive Gravity,
  Phys.\ Rev.\ D {\bf 88} (2013) 10,  106004,
  [arXiv:1308.4970 [hep-th]].

\bibitem{CL2}
L. Cheng, X. H. Ge and S. J. Sin, Anisotropic plasma with a chemical potential and scheme-independent instabilities, Phys. Lett. B {\bf 734} 116 (2014) [arXiv:1404.1994[hep-th]].
\bibitem{CL3}
L. Cheng, X. H. Ge and S. J. Sin, Anisotropic plasma at finite $U(1)$ chemical potential,  JHEP {\bf 07} 083 (2014)  [arXiv:1404.5027[hep-th]].
\bibitem{htcsc}A. Damaseelli, Z. Hussain, Z. X. Shen, Angle-resolved Photoemission studies of the cuprate superconductors. Rev. Mod. Phys. (2003) 75,473.
\bibitem{1504.05561}
  M. Baggioli, M. Goykhman,
  Holographic Polarons, the Metal-Insulator Transition and Massive Gravity,
  JHEP {\bf 07} (2015) 035,
  [arXiv:1504.05561 [hep-th]].
\bibitem{1411.1003}
  M. Baggioli, O. Pujolas,
  Holographic Polarons, the Metal-Insulator Transition and Massive Gravity,
  Phys. Rev. Lett. {\bf 114}, 251602 (2015),
  [arXiv:1411.1003 [hep-th]].

%
%

\end{thebibliography}
\end{document}